\documentclass[12pt]{article}
\usepackage{graphicx}
\newcommand{\eqn}[1]{(\ref{#1})}

\newcommand{\bra}[1]{\langle{#1}|}
\newcommand{\ket}[1]{|{#1}\rangle}
\newcommand{\braket}[2]{\langle{#1}|{#2}\rangle}

\def\beq{\begin{equation}}
\def\eeq{\end{equation}}
\def\beqa{\begin{eqnarray}}
\def\eeqa{\end{eqnarray}}

\newcommand{\EQ}{\begin{equation}}
\newcommand{\EN}{\end{equation}}
\newcommand{\bea}{\begin{eqnarray}}
\newcommand{\ena}{\end{eqnarray}}

\renewcommand{\a}{\alpha}

\newcommand{\NP}[1]{Nucl.\ Phys.\ {\bf #1}}
\newcommand{\PL}[1]{Phys.\ Lett.\ {\bf #1}}

\renewcommand{\thefootnote}{\fnsymbol{footnote}}
\def\one{{\hbox{ 1\kern-.8mm l}}}

\def\ii{{\rm i}}

\def\tr{{\rm tr\,}}

\newlength{\bredde}
\def\slash#1{\settowidth{\bredde}{$#1$}\ifmmode\,\raisebox{.15ex}{/}
\hspace*{-\bredde} #1\else$\,\raisebox{.15ex}{/}\hspace*{-\bredde} #1$\fi}
\newsavebox{\uuunit}
\sbox{\uuunit}
    {\setlength{\unitlength}{0.825em}
     \begin{picture}(0.6,0.7)
        \thinlines
        \put(0,0){\line(1,0){0.5}}
        \put(0.15,0){\line(0,1){0.7}}
        \put(0.35,0){\line(0,1){0.8}}
       \multiput(0.3,0.8)(-0.04,-0.02){12}{\rule{0.5pt}{0.5pt}}
     \end {picture}}

\newcommand {\Cbar}
    {\mathord{\setlength{\unitlength}{1em}
     \begin{picture}(0.6,0.7)(-0.1,0)
        \put(-0.1,0){\rm C}
        \thicklines
        \put(0.2,0.05){\line(0,1){0.55}}
     \end {picture}}}
\newsavebox{\zzzbar}
\sbox{\zzzbar}
  {\setlength{\unitlength}{0.9em}
  \begin{picture}(0.6,0.7)
  \thinlines
  \put(0,0){\line(1,0){0.6}}
  \put(0,0.75){\line(1,0){0.575}}
  \multiput(0,0)(0.0125,0.025){30}{\rule{0.3pt}{0.3pt}}
  \multiput(0.2,0)(0.0125,0.025){30}{\rule{0.3pt}{0.3pt}}
  \put(0,0.75){\line(0,-1){0.15}}
  \put(0.015,0.75){\line(0,-1){0.1}}
  \put(0.03,0.75){\line(0,-1){0.075}}
  \put(0.045,0.75){\line(0,-1){0.05}}
  \put(0.05,0.75){\line(0,-1){0.025}}
  \put(0.6,0){\line(0,1){0.15}}
  \put(0.585,0){\line(0,1){0.1}}
  \put(0.57,0){\line(0,1){0.075}}
  \put(0.555,0){\line(0,1){0.05}}
  \put(0.55,0){\line(0,1){0.025}}
  \end{picture}}
\newcommand{\Zbar}{\mathord{\!{\usebox{\zzzbar}}}}

\begin{document}
%\begin{titlepage}
\begin{flushright} KUL-TF-99/9 \\ DFTT-9/99 \\
February 1999
\end{flushright}
\vskip 0.6cm
\begin{center} 
{\Large \bf On D-branes in Type 0 String Theory\footnote{%
Work supported by the European Commission TMR programme ERBFMRX-CT96-0045.}}
\vskip 0.5cm  
{\bf   Marco Bill\'o$^{a,}$\footnote{%
E-mail: {\tt billo@to.infn.it}}, 
Ben Craps$^{b,}$\footnote{%
Aspirant FWO, Belgium; e-mail: 
{\tt Ben.Craps@fys.kuleuven.ac.be}} and 
Frederik Roose$^{b,}$\footnote{%
E-mail: {\tt Frederik.Roose@fys.kuleuven.ac.be}}} \\
\vskip .2cm
{\sl $^a$ Dipartimento di Fisica Teorica, Universit\`a di
Torino and}\\
{\sl I.N.F.N., Sezione di Torino, via P. Giuria 1, I-10125, Torino, Italy}
\vskip .2cm
{\sl $^b$ Instituut voor Theoretische Fysica}\\
{\sl Katholieke Universiteit Leuven, B-3001 Leuven, Belgium}
\end{center}
\vskip 0.2cm
\begin{abstract}
Using boundary states we derive the presence of (chiral) fermions on the 
intersection of type 0 D-branes. The corresponding anomalous couplings on 
the branes are then computed.
Furthermore, we discuss systems of branes at $\Cbar^2/\Zbar_n$
orbifold singularities. 
In particular, the massless spectrum on the branes is derived, and a boundary
state description is given. 
%The gauge coupling beta-function is shown to 
%vanish at large $N$.
\end{abstract}
%%%%%%%%%%%%%%%%%%%%%%%%%%%%%%
\renewcommand{\thefootnote}{\arabic{footnote}}
\setcounter{footnote}{0}
%%%%%%%%%%%%%%%%%%%%%%%%%%%%%%%
%%%%%%%%%%%%%%%%%%%%%%%%%%%%%%%%%%%%%%%%%%%%%%%%%%%%%%%%%%%%%%%%%%%%%%%%%%%%%%%%%%%%%%%%%%%%%%%%%%%%%%%%%
\section{Introduction}
Type 0 string theories \cite{dixharv} have recently attracted a lot of 
attention. The hope is that information on non-supersymmetric gauge theories 
can be extracted by embedding them in these non-supersymmetric
string theories \cite{polya, KT}. As in the adS/CFT correspondence for type II
string theory \cite{maldacena}, D-branes play a crucial role in this
conjectured string theory/gauge theory correspondence.   
In this paper, we will be concerned with the relation between type 0 D-branes
and the supersymmetric type II D-branes. We will show how certain results,
well established in the type II context, can be extended to the type 0 one. 

In the Neveu-Schwarz-Ramond formulation, type II string theories are obtained 
by imposing 
%the same (or the opposite in the R sector) 
independent GSO projections on the left and right moving part. 
This amounts to keeping the following (left,right) sectors: 
\beqa
{\rm IIB:}&&(NS+,NS+)\ ,~~(R+,R+)\ ,~~(R+,NS+)\ ,~~(NS+,R+)~;\nonumber\\
{\rm IIA:}&&(NS+,NS+)\ ,~~(R+,R-)\ ,~~(R+,NS+)\ ,~~(NS+,R-)~,\nonumber
\eeqa
where for instance R$\pm$ is the Ramond sector projected with 
$P_{\rm GSO}=(1 \pm (-)^F)/2$,
$F$ being the world-sheet fermion number.

There is another, equivalent choice for both theories 
(related to the first choice by a spacetime reflection):
\beqa
{\rm IIB':}&&(NS+,NS+)\ ,~~(R-,R-)\ ,~~(R-,NS+)\ ,~~(NS+,R-)~;\nonumber\\
{\rm IIA':}&&(NS+,NS+)\ ,~~(R-,R+)\ ,~~(R-,NS+)\ ,~~(NS+,R+)~.\nonumber
\eeqa
For the massless R-R sector the difference between the primed and 
unprimed theories shows up in opposite chiralities of the bi-spinor 
containing the R-R field strengths. This implies a sign difference in the
Poincar\'e duality relations among these field strengths, 
resulting in a selfdual five-form field strength 
in IIB and an antiselfdual one in IIB', for instance.

The type 0 string theories contain instead the following sectors:
\beqa
{\rm 0B:}&&(NS+,NS+)\ ,~~(NS-,NS-)\ ,~~(R+,R+)\ ,~~(R-,R-)~; \nonumber\\
{\rm 0A:}&&(NS+,NS+)\ ,~~(NS-,NS-)\ ,~~(R+,R-)\ ,~~(R-,R+)~.\nonumber
\eeqa
These theories do not contain bulk spacetime fermions, which would have to
come from ``mixed''
(R,NS) sectors.\footnote{However, fermions will occur when D-branes are
introduced \cite{gaberdiel}. This will be crucial for our results.}    
The inclusion of the NS-NS sectors with odd fermion numbers means that the 
closed string tachyon is not projected out. 
The third difference with type II theories is that the R-R spectrum is 
doubled: the R-R potentials of the primed and unprimed type II theories 
are combined, resulting for instance in an unconstrained five-form
field strength in type 0B. 

Because of the doubling of the R-R spectrum, there are in type 0B two kinds
of D3-branes, named electric and magnetic in constrast 
to the selfdual D3-brane of type IIB. 
The recent literature has developed in two main directions. First, the gauge theory on a superposition 
of a large number of electric branes was studied, leading to non-supersymmetric, non-conformal, 
tachyon-free gauge theories \cite{polya, 9811035, minahan}. 
Second, equal numbers of electric and magnetic D3-branes were superposed, 
giving a non-supersymme\-tric large $N$ conformal field
theory \cite{KT, tseytzar, nekrasov}.

In this paper we will be mostly interested in the second situation. 
A superposition of equally many electric and magnetic branes is reminiscent 
of the type II branes (see Section 2), which have been well studied. It will
turn out that a lot of results can be transferred to type 0 almost without 
effort. For instance (Section 3), chiral fermions are present on certain 
intersections of electric and magnetic branes, leading to anomaly inflow on
the intersection via anomalous D-brane couplings. 
These, in turn, lead to the creation of a string when certain D-branes cross 
each other. 

In Ref. \cite{DM}, the configuration of many type IIB D3-branes sitting at 
a $\Cbar^2/\Zbar_n$ orbifold singularity was considered, and the spectrum
of the corresponding supersymmetric world-volume theory was described. 
In Section 4, we consider the analogous type 0 situation. We will
derive the non-supersymmetric massless spectrum of equally many electric 
and magnetic type 0  branes at a  $\Cbar^2/\Zbar_n$ orbifold  singularity. 
The theory can be seen as an orbifold of the large $N$ conformal field
theory \cite{KT} of equally many electric and magnetic D3-branes
in flat space. According to an argument of Ref. \cite{nekrasov},
it is still conformal in the large $N$ limit. We have checked this 
by computing that  the gauge coupling beta function
vanishes, in that limit, up to two-loop order. 

In many respects type 0 string theory is the result of a clever way 
of breaking supersymmetry. In taking the orbifold from type II to type 0 
the resulting theory, although non-supersymmetric, still
exhibits features reminiscent of the parent theory. 
{}From the strong resemblances between type II and type 0 string theories, 
one may hope to draw more conclusions about non-supersymmetric 
theories by reinterpreting old results 
from supersymmetric string theory in this new context.

%A second point of interest will be an orbifold of the large $N$ CFT of 
%Ref. \cite{KT}. The massless spectrum of
%equally many electric and magnetic type 0 branes at an $A_n$ orbifold 
%singularity will be derived and a boundary state description of this derivation will be given. 
%The large $N$ conformal invariance of the resulting gauge theory will be
%illustrated by a two loop computation of the gauge coupling beta function.

%We have organized the remainder of this paper as follows. In Section 2 D-branes in type 0 theories are
%quickly reviewed. Section 3 contains our analysis of anomaly inflow on type 0 brane intersections and
%related issues. Section 4 discusses the spectrum of the orbifold CFT, its large $N$ conformal
%invariance and the boundary state description. Our conclusions are collected in Section 5.

%%%%%%%%%%%%%%%%%%%%%%%%%%%%%%%%%%%%%%%%%%%%%%%%%%%%%%%%%%%%%%%%%%%%%%%%%%%%%%%%%%%%%%%%%%%%%%%%%%%%%%%%%
\section{D-branes in type 0 string theory}
\label{s:type0D}
D-branes in type 0 theories have been discussed in Refs \cite{gaberdiel} and 
\cite{9811035}. In this section we mainly review some of their results. 

As discussed in the introduction, the spectrum of type 0 theories contains two 
($p+1$)-form R-R potentials for each even (0A) or odd (0B) $p$. We will
denote these by $C_{p+1}$ and $C'_{p+1}$, referring to the unprimed and 
primed type II theories mentioned above. For our purposes, more convenient 
%way to label these fields is 
combinations are
\beq \label{C+-}
(C_{p+1})_\pm=\frac{1}{\sqrt{2}}(C_{p+1}\pm C'_{p+1})~.
\eeq 
For $p=3$ these are the electric ($+$) and magnetic ($-$) potentials 
\cite{9811035}. We will adopt this terminology also for other values of $p$. 

There are four types of ``elementary'' D-branes for each $p$: an electric and a 
magnetic one ({\it i.e.}, charged under $(C_{p+1})_\pm$), and the
corresponding antibranes. 
In  Ref. \cite{9811035} the interaction energy of two identical parallel 
($p+1$)-branes was derived by computing the relevant
%\footnote{%
%For instance, for like branes the contribution of the
%open string Ramond sector is absent: consistently with the absence of
%space-time fermions in the bulk, there are no worldvolume fermions.}
cylinder diagram in the open string channel, analogously to the Polchinski 
computation~\cite{Pol} in type II. Isolating, via modular transformation, the 
contributions due to the exchange of long-range fields in the closed string 
channel, it is found on the one hand that the tension of these branes is a 
factor $\sqrt 2$ smaller than for type II  branes. 
On the other hand, the R-R repulsive force between two like branes  
has the double strength of the graviton-dilaton attraction \cite{9811035}; 
thus the type 0 branes couple to the corresponding R-R potentials 
$(C_{p+1})_\pm$ with the same charge as the branes in type II couple to the  
potential $C_{p+1}$.

The cylinder diagram between two D-branes in type II can also be considered
as a tree-level diagram in which a closed string propagates
between two ``boundary states''. 
A boundary state is a particular BRST invariant closed string state
that describes the emission of  a closed string from a D-brane. It
satisfies conditions\footnote{%
We refer to Ref. \cite{billo9802} 
% and \cite{benfred} 
for explicit expressions, conventions and normalizations 
for  boundary states; see also later Section 4.1.}
that correspond to the boundary conditions for open strings ending on 
the D-brane. 
In particular, for the fermionic fields $\psi^\mu$ the boundary state 
$\ket{B,\eta}_{\rm NS,R}$, which depends on the sector, R or NS, 
%of $\psi^\mu$ and $\tilde\psi^\mu$ 
and on an additional sign $\eta=\pm$, satisfies 
%For instance, the open string Neumann (Dirichlet) b.c.'s 
%translate into the conditions $\partial_\tau X^\alpha |_{\tau=0}\ket{B} =  
%X^i|_{\tau=0}\ket{B}=0$ for the closed string fields 
%$X^\alpha$ and $X^i$ in the worldvolume and transverse directions respectively.
\beq
(\psi^\mu-\ii\eta S^\mu_{~\nu}\tilde\psi^\nu)\ket{B,\eta}_{\rm NS,R}=0~,
\eeq
where $S^{\mu}_{~\nu}$ is diagonal, with entries $1$ in the worldvolume
and $-1$ in the transverse directions. 
In type II theories, the GSO projection requires a mixture of the two choices
$\eta=\pm 1$. 
Indeed, starting, {\em e.g.,}  with $\eta=+1$, one finds that
the type II boundary state 
$\ket B$ $=P_{\rm GSO}\ket{B,+}_{\rm NS}$ 
$\oplus P_{\rm GSO}\ket{B,+}_{\rm R}$
is
\begin{equation}
 \label{am1}
 \ket B = 
 {1\over 2}\left((\ket{B,+}_{\rm NS} - \ket{B,-}_{\rm NS})\oplus 
 (\ket{B,+}_{\rm R} +\ket{B,-}_{\rm R})\right)~.
\end{equation} 

We remark that in the case of type 0 D-branes, 
the various cylinder amplitudes between an electric (or magnetic) D-brane 
and  an electric (or magnetic) one can simply be reproduced using the 
following unprojected (and differently normalized) boundary states, whose sum is 
$\sqrt{2}$ times the type II one: 
\beq \label{B+-}
\ket{B,\pm}=\frac{1}{\sqrt 2}\left(\pm\ket{B,\pm}_{\rm NS}\oplus 
\ket{B,\pm}_{\rm R}\right)~.
\eeq
Here $\ket{B,+}$ represents an electric brane and $\ket{B,-}$ a magnetic one.
%\beq
%|B,->=\frac{1}{\sqrt 2}(-|B,->_{\rm NS}+ |B,->_{\rm R})
%\eeq
%for a magnetic one. Here we are using the conventions and normalizations of 
% 
%which we will use troughout
%the paper.\footnote{Note that, in these conventions, the {\it bra} of an electric brane is given by 
%$<B,-|$ and its {\it ket} by $|B,+>$.} The $\pm$ refers to the
%periodicity conditions on the fermions at the boundary described by the boundary state, {\it e.g.},
%\beq
%(\psi_m-\ii\eta S\cdot\tilde\psi_{-m})|B,\eta>=0~.
%\eeq

As a cross-check, with these boundary states it is easy to compute the 
one-point function on the disc of a R-R potential (as in Ref. \cite{dv9707}).
Denoting by $\ket{C_{p+1}}_\pm$ the state corresponding to the 
$(C_{p+1})_\pm$ potentials, the one-point function describing its coupling
to a type 0 D$p$-brane will be
$\braket{B,\pm}{C_{p+1}}_\pm$. This gives indeed  
the same charge as in type II, where we would compute
$\braket{B}{C_{p+1}}$, as we can see from Eqs 
(\ref{C+-},\ref{am1},\ref{B+-}). 
% (the factors
%$\frac{1}{\sqrt 2}$ in Eqs \eqn{C+-} and \eqn{B+-} are compensated by the factor $2$ one gets from not having to do
%the type II GSO projection).
%%%%%%%%%%%%%%%%%%%%%%%%%%%%%%%%%%%%%%%%%%%%%%%%%%%%%%%%%%%%%%%%%%%%%%%%%%%%%%%%%%%%%%%%%%%%%%%%%%%%%%%%%%%%%
\section{Anomaly inflow and Wess-Zumino action}
The D-branes described in the previous section showed many similarities
to their type II cousins. In this section
we will push the analogy further to include the whole Wess-Zumino action, 
{\it i.e.}, all the anomalous D-brane
couplings\footnote{and the non-anomalous ones found in Ref. \cite{normal}}. 
%We will do this in two ways, first by repeating the anomaly inflow argument 
%of Refs \cite{GHM, CY}, and second by repeating the explicit computation of 
%scattering amplitudes of closed string fields on the disc
%\cite{benfred, stefanski, normal} 
%(see Ref. \cite{MSS} for an alternative string computation of these terms).

The open strings stretching between two like branes are bosons, just like the 
bulk fields of type 0. However, fermions appear from strings between an 
electric and a magnetic brane \cite{gaberdiel}. Thus one could
wonder whether there are chiral fermions on the intersection of an electric 
and a magnetic brane. Consider such an orthogonal intersection with no overall 
transverse directions. If the dimension of the intersection is
two or six, a cylinder computation reveals that there are precisely enough 
fermionic degrees of freedom on the intersection 
%stretching between the branes 
to form one chiral fermion.

In type II string theory, the analogous computation shows that chiral fermions 
are present on two or six dimensional intersections of two orthogonal branes 
with no overall transverse directions. That observation has
had far reaching consequences. Namely, the presence of chiral fermions has been shown to
lead to gauge and gravitational anomalies on those intersections of D-branes 
%(sometimes called I-branes)
\cite{GHM}. In a consistent theory, such anomalies should be cancelled by anomaly inflow \cite{CH}. In the
present case, the anomaly inflow is  provided by the anomalous D-brane couplings in the Wess-Zumino part of
the D-brane action \cite{GHM,CY}. These anomalous couplings have an anomalous variation localized on the
intersections with other branes.     

To sketch how this anomaly inflow comes about, let us focus on the case of two 
type IIB D5-branes (to be denoted by D5 and D5') intersecting on a string. 
The Wess-Zumino action on  D5 contains a term of the form
$\int_{\rm D5}C_2\wedge Y_4$, where $C_2$ is the R-R two-form potential 
and $Y_4$ a certain four-form involving the field strength of gauge field on 
D5 and the curvature two-forms of the tangent and normal bundles of D5. To be precise, one
should replace this term by $\int_{\rm D5}H_3\wedge \omega_3$, with 
$Y_4=d\omega_3$ and $H_3$ the complete
gauge-invariant field strength of $C_2$ (which generically differs from $dC_2$). With the additional
information that the gauge variation of the ``Chern-Simons''
form $\omega_3$ is given by $\delta\omega_3=dI_2$ for some two-form $I_2$,
it is easy to see that the anomalous term on D5 will have a variation localized on the intersection with D5':
\beq	
\delta \int_{{\rm D}5}\,H_3\wedge \omega_3=\int_{{\rm D}5}\,dH_3\wedge I_2=
\int_{{\rm D}5}\,d*H_7 \wedge I_2=
\int_{{\rm D}5}\,\delta_{{\rm D}5'}\wedge I_2~.
\eeq 

A careful analysis of all the anomalies \cite{CY} shows that the anomalous 
part of the D$p$-brane action is given, in terms of the formal sum $C$
of the various R-R forms, by
\beq
S_{\rm WZ}=\frac{T_p}{\kappa}\int_{p+1} C\wedge e^{2\pi\a '\,F+B}\wedge
\sqrt{\hat{A}(R_T)/\hat{A}(R_N)}~~.
\eeq
Here $T_p/\kappa$ denotes the D$p$-brane tension, $F$ the gauge field on 
the brane and $B$ the NS-NS two-form.
Further, 
%$\hat{A}(R_T)$ and $\hat{A}(R_N)$ denote the A-roof genus of the
%tangent and normal bundles of the D-brane world-volume, respectively.
%The A-roof genus is a polynomial in the curvature, the precise form of which
%will not be needed in this paper.
$R_T$ and $R_N$ are the curvatures of the tangent and normal 
bundles of the D-brane world-volume, and $\hat{A}$ denotes the A-roof genus:
\beqa
\label{Aroof}
\sqrt{\frac{\hat{A}(R_T)}{\hat{A}(R_N)}}&=&1+\frac{(4\pi^2\a ')^2}{384\pi^2}(\tr R_T^2-\tr R_N^2)
+\frac{(4\pi^2\a ')^4}{294912\pi^4}(\tr R_T^2-\tr R_N^2)^2\nonumber\\&&+
\frac{(4\pi^2\a ')^4}{184320\pi^4} (\tr R_T^4-\tr R_N^4)+\ldots~.
\eeqa
This action has been checked  in Refs \cite{benfred, MSS, stefanski, normal}
by computing various string scattering amplitudes.

These anomalous D-brane couplings have various applications. To mention just one, using T-duality it has been 
argued \cite{BDG} that they imply the creation of a fundamental string whenever certain type II D-branes cross
each other. This string creation process is dual to the Hanany-Witten effect \cite{hanany}.

Let us now return to type 0 string theory. As stated above, here chiral fermions live on intersections of 
electric and magnetic type 0 D-branes.  
The associated gauge and gravitational anomalies on such intersections 
match the ones for type II D-branes. To cancel them, the minimal coupling of a
D$p$-brane to a ($p+1$)-form R-R potential should be extended to the following 
Wess-Zumino action:   
\beq
\label{WZaction}
S_{\rm WZ}=\frac{T_p}{\kappa}\int_{p+1}(C)_\pm\wedge 
{\rm e}^{2\pi\a '\,F+B}\wedge
\sqrt{\hat{A}(R_T)/\hat{A}(R_N)}~.
\eeq
The $\pm$ in Eq. \eqn{WZaction} distinguishes between electric and magnetic 
branes. Note that $T_p/\kappa$ denotes the tension
of a type II D$p$-brane, which is $\sqrt 2$ times the type 0 D$p$-brane tension.

The argument that the variation of this action\footnote{Again, to be precise, 
as in type II \cite{GHM, CY} one should use an action expressed in 
terms of the R-R field strengths instead of the potentials, 
which is different from Eq. \eqn{WZaction}.}
cancels the anomaly on the intersection is a copy of
the one described above in the type II case, apart from one slight
subtlety. For definiteness, consider the intersection of an electric and a 
magnetic D$5$-brane on a string.
Varying the electric D$5$-brane action (exhibiting the $(C_2)_+$ potential, 
or rather, its field strength $(H_3)_+$), one finds that the variation 
is localized on the intersection of the electric  D$5$-brane with
branes charged magnetically under the $(H_3)_+$ field strength. 
Using Eq. (\ref{C+-}), the different behaviour under Poincar\'e duality of the R-R field strengths 
of type II and II' shows that these are precisely the branes  
carrying (electric) $(H_7)_-$ charge, {\it i.e.}, 
what we called the magnetic D$5$-branes. 
Schematically,
\beq
\delta \int_{{\rm D}5_+}\,(H_3)_+\wedge \omega_3=\int_{{\rm D}5_+}\,d(H_3)_+
\wedge I_2= \int_{{\rm D}5_+}\,d*(H_7)_-\wedge I_2=
\int_{{\rm D}5_+}\,\delta_{{\rm D}5_-}\wedge I_2~.
\eeq
A completely analogous discussion goes through for the
variation of the magnetic D$5$-brane action. 

This anomaly inflow argument fixes (the anomalous part of) the Wess-Zumino action, displayed in Eq. \eqn{WZaction}.
The presence of these terms (and of a similar non-anomalous one \cite{normal}) can be checked by a disc
computation, as in type II \cite{benfred, stefanski, normal}. 
In fact, up to the cancelling factors mentioned at the end of the previous 
section, the computation is precisely the same, confirming the form of the
action \eqn{WZaction}.

Assuming T-duality to hold between type 0A/B, the arguments of Ref. \cite{BDG} lead to the creation of a fundamental
string when certain electric and magnetic branes cross each other. In type II this is linked to the Hanany-Witten
effect \cite{hanany} by a chain of dualities. However, this chain involves S-duality, of which we do not know a
type 0 analogue.
%%%%%%%%%%%%%%%%%%%%%%%%%%%%%%%%%%%%%%%%%%%%%%%%%%%%%%%%%%%%%%%%%%%%%%%%%%%%%%%%%%%%%%%%%%%%%%%%%%%%%%%%%%%%%%
\section{Type 0 branes at an orbifold fixed point} \label{s:fields}
In this section we want to find the open string spectrum on a system of 
$N$ electric and $N$ magnetic type 0 D-branes at an orbifold fixed point. 
As in section 2 the type 0 situation closely 
parallels the well-understood type II case. 
%Therefore we have the feeling that a brief review
%of the type II construction may be useful. To go to the type 0 case is then a one-step 
%procedure which is outlined next. Finally, a closed string "derivation" is given, which along
%the lines of section 2 presents
%supportive evidence for the open string massless spectrum. 

\begin{figure}
\centering
\includegraphics[width=10cm]{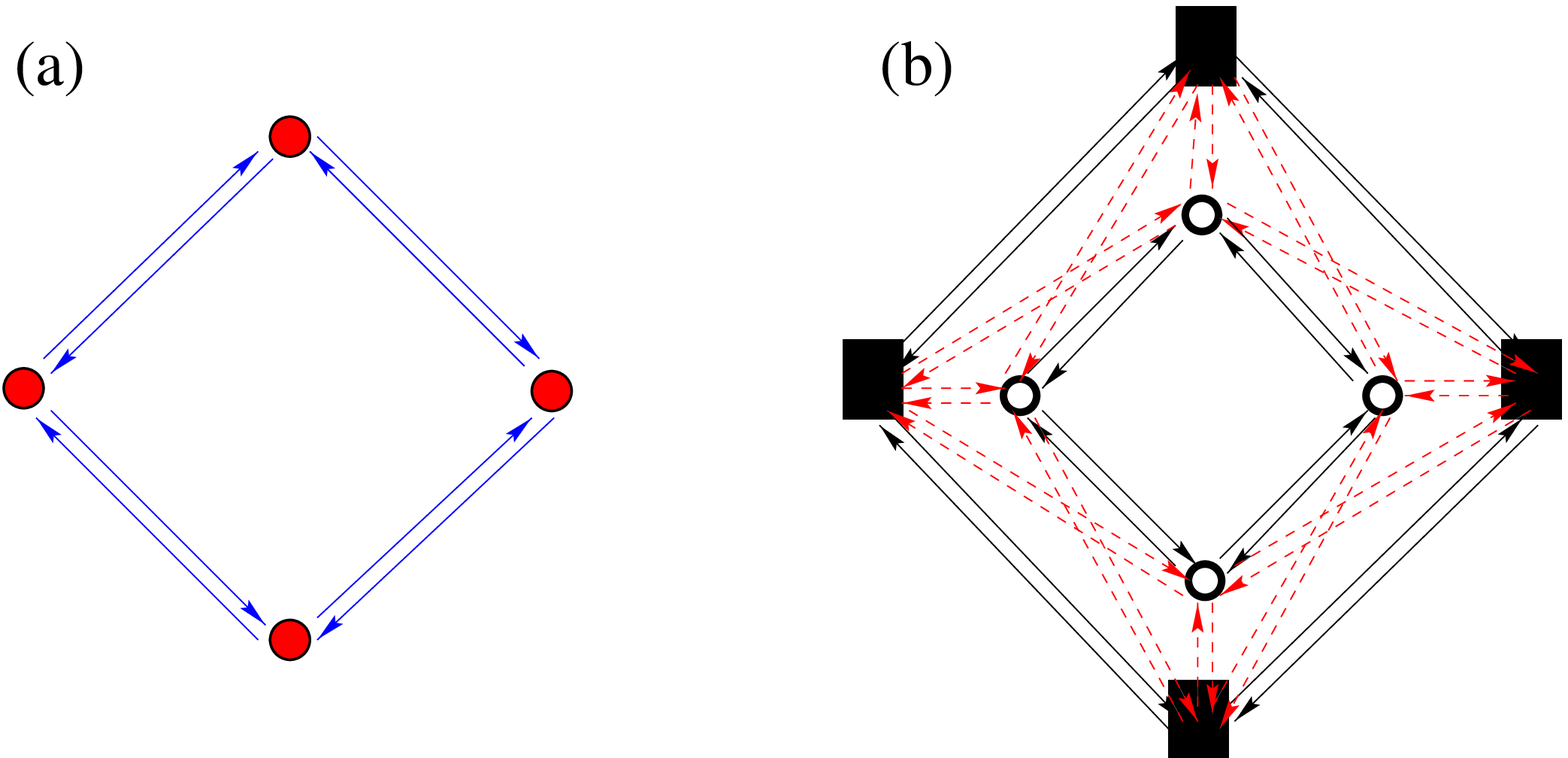}
\caption{(a) \textsl{Type II quiver diagram for the $\Zbar_4$ orbifold. (b)
Its type 0 counterpart: the links drawn with solid lines correspond to
bosons, the other ones to fermions.}}
\label{fig1}
\end{figure}

In type IIB theory, D5-branes sitting at the singularity of a $\Cbar^2/\Gamma$ orbifold are 
{\em defined} by D5-brane configurations on the covering space $\Cbar^2$. 
Here and below $\Gamma$ denotes a discrete subgroup of SU$(2)$. 
The definition is such that the action of $\Gamma$ is extended to include 
the (open string) Chan-Paton factors. Let us take $\Gamma =
\Zbar_n$ for concreteness. To describe $N$ D5-branes on this orbifold, 
one starts with $nN$ D5-branes on $\Cbar^2$, reflecting that if a point 
$P$ is an allowed open string endpoint, then so are its image points under 
the orbifold group. The open string sector thus consist of
a vector $A_\mu$ ($\mu=0,\ldots 5)$ and 4 scalars $X^I$ ($I=6,\ldots 9)$
in the adjoint of SU$(nN)$. The spacetime action of $\Zbar_n$ is diagonal on 
the complex scalars $X= X^6 + \ii X^7$ and  $Y= X^8 + \ii X^9$:
its generator sends $(X,Y)$ into $(\omega X,\omega^{-1}Y)$, where
$\omega^n=1$.
%
%on $X^I$ is represented by a matrix 
%$R^J_{\ I}$ acting on $X^I$, while the vector is
%left invariant. 
This geometrical action is then supplemented by an action 
on the Chan-Paton factors. First divide the $nN\times nN$ hermitian Chan-Paton factor
into $N\times N$ blocks, say $\lambda_{ij}, i,j=1\ldots n$. In an appropriate basis the
generator $g$ of $\Zbar_n$ is taken to act as
$g(\lambda_{ij}) = \omega^{i-j}\lambda_{ij}$, corresponding to the  regular 
action of $\Zbar_n$ on both indices.
%Likewise, the $X^I$ are assembled into
%two complex scalars $X,\bar X$, on which the spacetime orbifold rotation acts diagonally as
%multiplication by $\xi, \xi^{-1}$ respectively. 
As usual in open string theory the orbifold
projection is then performed so as to retain only $\Zbar_n$ invariant states. 
After 
%trivial
dimensional reduction from $d=6$ to $d=4$, one finds the following 
${\cal N}=2$, $d=4$ supermultiplets. There are $n$ vector multiplets
for different SU$(N)$ groups, and $n$ hypermultiplets that transform each in
the bifundamental of a couple of SU$(N)$ factors, as encoded in a type II
``quiver diagram'' \cite{DM}, see Fig. \ref{fig1}(a): each dot represents
a vector multiplet in one of the SU$(N)$ factors and each link a
bifundamental half hypermultiplet.
%%%%%%%%%%%%%%%%%%%%%%%%%
%%%%%%%%%%%%%%%%%%%%%%%%%%%

Next, let us consider the analogous situation for $N$ electric 
and $N$ magnetic type~0 D-branes. As discussed in Ref. \cite{KT} for the 
flat space case, one starts with $2N$ type II D-branes, 
derives the type II open string spectrum next, and finally
performs a $(-)^{F_{\rm s}}\,{\cal I}$ projection to find the type 0 spectrum.  
$F_{\rm s}$ is the space-time fermion number
and ${\cal I}$ acts as conjugation by $\sigma_3$ in the space of 
$N\times N$ blocks obtained by separating each group of 
$2N$ type II branes into $N$ electric and $N$ magnetic type 0 ones. 
The type 0 field content is as follows. 
The massless bosons are:
\begin{itemize}
\item 1 vector and 2 real scalars in the adjoint of each of the $n$ 
 ${\rm SU}(N)\times {\rm SU}(N)$ factors of the gauge group;
\item $n$ complex scalars, each in the representation
$((N,1),(\bar N,1))+ ((\bar N,1),(N,1))$ of two ${\rm SU}(N)\times {\rm SU}(N)$   
factors, as encoded in a type 0 quiver diagram,
wich we will introduce shortly,
see Fig. \ref{fig1}(b).
\end{itemize}
The massless fermionic content consists of: 
\begin{itemize}
\item 2 Weyl fermions in the $(N,\bar N)+(\bar N,N)$ 
of each ${\rm SU}(N)\times {\rm SU}(N)$ group factor;
\item $n$  Weyl fermions, each in the 
$((N,1),(1,\bar N)) + ((\bar N,1),(1, N))$ of a couple
of gauge factors as encoded in the type 0 diagram of Fig. \ref{fig1}(b). 
\end{itemize}
The type 0 quiver diagram that summarizes the information about the spectrum
requires only a slight modification of the quiver rules in Ref. \cite{DM}.
For instance, in our $\Zbar_n$ case, one draws $n$ pairs 
consisting of a filled  and an empty dot, 
representing the groups of electric and magnetic branes, respectively; 
one then draws oriented links between any two dots inside one pair or in
neighbouring pairs. 
The spectrum can be read off using the following rules:  
\begin{enumerate}
\item 
every dot represents the bosonic content of an SU$(N)$ vector multiplet;
\item
every link connecting two like dots corresponds to spacetime bosons;
\item
every link between an empty and a filled dot corresponds to spacetime fermions;
\item
a link represents the bosonic or fermionic truncation of either a vectormultiplet (in the case of
links within a pair) or
half a hypermultiplet (for links from one pair to a neighbouring one), 
transforming in the bifundamental of the gauge groups 
corresponding to the dots it connects. 
\end{enumerate}
%The origin of these rules is self-explanatory from Ref. \cite{DM} and the orbifolding to type
%0.
%%%%%%%%%%%%%%%%%%%%%%%%%%%%%%
\subsection{Closed string description}
%As of D-branes in flat space, also of D-branes at an orbifold singularity
%it is possible to give a description in terms of boundary states.
As we saw in Section 2, in terms of unprojected R and NS boundary states
one can with equal ease describe type II and type 0 D-branes.
%In this language, 
One can thus directly compute in the closed string channel
the one-loop open string free energy that encodes the free spectrum of 
the gauge theory on the branes, both for type II and for type 0.
%This gives a check of the type 0 spectrum derived in the previous section.
We give now in some detail the closed-string description of the configuration
treated above, namely D-branes sitting 
at an orbifold singularity (see also \cite{BI}); in particular, we will show 
how to incorporate the  non-trivial action of the 
orbifold group on the Chan-Paton factors. As far as we know, this prescription
is new, also in the type IIB context.
%In this subsection, we briefly discuss the unprojected boundary state for 
%a D-brane at an orbifold singularity (see \cite{BI}), which enables us to 
%compute the cylinder amplitude between two D-branes in the 
%closed string channel. We further discuss how to include 
%in this computation the effect of a
%non-trivial action of the orbifold group on the Chan-Paton factors.    

Let us thus consider D5-branes on $\Cbar^2/\Zbar_n$. 
%Given the action
%of the $\Zbar_n$ orbifold group on the $\Cbar^2$ coordinates described in 
%Section 4, t
The theory possesses 
$n$ closed string sectors, distinguished by the periodicity properties 
of the fields (see, {\it e.g.}, Ref. \cite{dixon}):
\begin{equation}
\label{twist}
\{X,Y,\psi_X,\psi_Y\}(\tau,\sigma+2\pi)= 
\{\omega^l X,\omega^{-l} Y,\omega^{l}\psi_X,\omega^{-l}\psi_Y\}(\tau,\sigma)~.
\end{equation}
Generically, the modings of the left and right moving oscillators,  
$\{\alpha_\nu,\beta_\lambda,\psi_\nu,\chi_\lambda\}$ and 
$\{\tilde\alpha_{\tilde\nu},\tilde\beta_{\tilde\lambda},
\tilde\psi_{\tilde\nu},\tilde\chi_{\tilde\lambda}\}$,
are shifted from integer values: 
$\nu, \tilde\lambda\in \Zbar + l/n$ and $\lambda, \tilde\nu\in \Zbar - l/n$
(plus of course an additional shift of 1/2 for the fermionic oscillators 
in the NS sector).
In the following we give explicit expressions for the $X$ and $\psi_X$ parts
only.
%,those for $Y$ and $\psi_Y$ being obtained simply by the replacement $\omega\to
%\omega^{-1}$.

A D5-brane localized at the  orbifold fixed point 
%in the transverse directions 
is represented in the $l$-th  twisted sector by a boundary state 
that
satisfies\footnote{%
Here and below we display only those parts of the conditions on the boundary
state, and of the  boundary state itself, that
differ from the usual flat space expressions given, {\em e.g.}, in
\cite{billo9802}. The parts containing the fields in the Neumann directions,
the ghost and superghosts are unaffected.}
\begin{equation}
\label{m4}
(X^{(l)}\,+\,\tilde X^{(l)})|_{\tau=0}\ket{B;l}=0~,
\hskip 0.3cm
(\psi^{(l)}\, +\, \ii \eta\, \tilde\psi^{(l)})|_{\tau=0}
\ket{B;l}=0~,
\end{equation}
plus conjugate relations and plus appropriate counterparts on $Y,\bar Y$ and
$\psi_Y,\bar\psi_Y$. 

The solution to these conditions in terms of oscillators\footnote{%
In twisted sectors, 
$l\not= 0$, we must take into account the fact that the only $\psi$ Ramond 
0-modes are those in the 6 Neumann directions, and modify the zero-mode part
of the R boundary states appropriately. In the NS sector, 
additional $\psi$ zero-modes must be taken into account
in the orbifold directions when $l/n=1/2$.}
is, writing $\ket{B;l}$ as the product 
$\ket{B_{\rm b};l}\otimes\ket{B_{\rm f},\eta;l}_{\rm R,NS}$
of a bosonic and a fermionic part,
\begin{eqnarray}
\label{m5bis}
\ket{B_{\rm b};l} & = & 
\exp\Bigl(\sum_\nu \,{\bar\alpha_{-\nu}\tilde\alpha_{-\nu}\over\nu} +
\,\sum_{\bar \nu} 
{\alpha_{-\bar\nu}\tilde{\bar\alpha}_{-\bar\nu}\over\bar\nu}
\Bigr)\ket{0}~,
\nonumber\\
\ket{B_{\rm f},\eta;l}_{\rm R,NS} & = &
\exp\Bigl(-\ii\eta\,\sum_\nu \,
\bar\psi_{-\nu}\tilde\psi_{-\nu} -\ii\eta\,
\sum_{\bar \nu} \psi_{-\bar\nu}\tilde{\bar\psi}_{-\bar\nu}
\Bigr)\ket{0}~,
\end{eqnarray}
where the modings are as indicated below Eq. (\ref{twist}) and, as usual, 
the fermionic part of the boundary state depends on the additional 
sign $\eta$.
% and differs in the R-R or NS-NS sector. 

Suppose now to have $nN$ D-branes, which for our purposes
we can represent
%\footnote{%
%Only for certain purposes, as for instance our present one, 
%the computation in the closed string channel of the open string free energy at 
%zero background:
%we are of course not able to describe the true non-abelian boundary states.} 
by $n N$ boundary states, labeled by ``Chan-Paton'' indices.
In the case we are interested in, we group the D5-branes 
into $n$ bunches 
%of $2N$ each, 
and use a Chan-Paton composite 
label $i,A$, with $i=0,\ldots, n$ and $A=1,\ldots, N$. 
In the type 0 context, we start with $A=1,\ldots, 2N$ and we split it
into $a=1,\ldots, N$, for the $N$ electric branes and $\bar a=1,\ldots, N$ 
for the magnetic ones.

The cylinder amplitude between any two given D-branes, 
{\em i.e.}, for fixed Chan-Paton indices, receives contributions
from all the twisted sectors. In summing the twisted sectors,
an ambiguity shows up: for each Chan-Paton label we can independently
decide how to weigh the twisted sectors. 
This is the closed string counterpart of the freedom
one has in the open string picture to introduce a non-trivial action 
of the orbifold group on the adjoint Chan-Paton factor. 
%A particular 
%non-trivial action, corresponding to the regular representation, 
%was considered in Ref. \cite{DM}; for instance in the $\Cbar^2/\Zbar_n$
%case the generator acts as $(m,a;m',a') \to \omega^{m-m'}(m,a;m',a')$
%in a convenient basis.
In the $\Cbar^2/\Zbar_n$ case, the regular action 
$(i,a;j,b) \to \omega^{i-j}(i,a;j,b)$ of the orbifold generator 
%chosen in the $\Cbar^2/\Zbar_n$ case in \cite{DM} and 
described at the beginning of Section 4
%\footnote{This regular action plays a crucial role
%in the argument of Refs \cite{nekrasov} that conformal invariance is preserved
%in the large-$N$ limit after orbifolding.} 
%In the closed string language, this 
is reproduced in the closed string language by defining the 
cylinder amplitude
to be\footnote{%
We fix the relative normalization of the boundary state in the various 
twisted sectors requiring that, for instance in the case of a single D-brane, 
the modular transformation of the closed string amplitude coincides with the 
free energy of open strings {\em projected} onto $\Zbar_n$-invariant states 
(see later).
Explicitly, the normalization in front of the boundary state is 
$T^{(l)}_5/(2\sqrt{n})$, with $T^0_5=\sqrt{\pi}(2\pi\sqrt{\alpha'})^{-2}$ 
being the usual D5-brane tension, while 
$T^{(l)}_5 =  \sqrt{\pi}8\pi^2\alpha'^{-2}\, 2\sin(\pi l/n) $ when $l\not=0$ 
%this is consistent with removing a factor of $(2\pi^2\alpha')^{1/4}$,
%which is naturally associated to each bosonic zero-mode in a D direction,
%when this zero mode is absent
.}
\begin{equation}
\label{m7}
({\cal A}^{ij})^{ab} = \sum_{l=0}^{n-1}\omega^{l(i-j)}\, 
 \bra{B,\eta';l;i,a}D\ket{B,\eta;l;j,b}~,
\end{equation}
where $D$ is the closed string propagator and of course the R or NS fermions
must be distinguished.
Explicitly, the amplitude (\ref{m7}) turns out to be
%, in matrix notation,
%The computation of the amplitude (\ref{m7}) is rather straightforward, 
%and the final result is
\begin{equation}
\label{m8}
({\cal A}_\gamma^{ij})^{ab} = \ii V_6 (-)^{\delta_{\gamma,2}} (8\pi^2\alpha')^{-3}
\int_0^{-\ii\infty} d\tau\,{1\over n}\sum_{l=0}^{n-1}\omega^{l(i-j)}
%4\sin^2{\pi l\over n} 
\,c_l\,
\left({f_\gamma(\tau)\over f_1(\tau)}\right)^4
\left({\theta_\gamma(\tau l/n|\tau)\over\theta_1(\tau l/n|\tau)}
\right)^2~,
\end{equation} 
where $V_6$ is the 5-brane volume, $c_0 = 1$ and $c_l=4 \sin^2(\pi l/n)$
for $l>0$  and we have $\gamma=2$ for the R-R sector with
$\eta'\eta=-1$, $\gamma=3$ for the NS-NS sector with $\eta'\eta=-1$ and
$\gamma=4$ for the NS-NS sector with $\eta'\eta=1$. The R-R amplitude with 
$\eta'\eta=1$ vanishes. In Eq. (\ref{m8}) we have used also the 
%standard
%by now customary
expressions $f_\gamma(\tau)$ of Ref. \cite{polcai}.
%In the case $l=0$ the integrand in Eq. (\ref{m8}) is actually
%rewritten as $\tau^{-2}(f_A/f_1)^8$, the usual expression for
%parallel D5-branes at zero transverse distance.  

The modular transformation of the amplitude (\ref{m8})
allows one to recognize it as a 1-loop open string free energy. 
In terms of $\tau' = -1/\tau$, $({\cal A}_\gamma^{ij})^{ab}$ is rewritten as
\begin{equation}
\label{m9}
%{\cal A}_A^{ij}= 
({\tilde{\cal A}}_{\gamma'}^{ij})^{ab}= -\ii V_6 (-)^{\delta_{\gamma',4}}
\int_0^{\ii\infty} d\tau'(8\pi^2\alpha')^{-3}
{1\over n}\sum_{l=0}^{n-1}\omega^{l(i-j)}
%4\sin^2{\pi l\over n} 
\, c_l \,
\left({f_{\gamma'}(\tau')\over f_1(\tau')}\right)^4
\left({\theta_{\gamma'}(l/n|\tau')\over\theta_1(l/n|\tau')}
\right)^2~,
\end{equation}   
with $2'=4$, $3'=3$ and $4'=2$.
The full amplitude ${\tilde{\cal A}}_{\gamma'}$, obtained summing 
%${\cal A}_{i'}^{m,a;m',a'}$ 
over all adjoint Chan-Paton indices, 
coincides with the 1-loop free energy for the $\Zbar_n$-invariant states of the 
open strings attached to the
D5-branes:
%, to which only $\Zbar_n$-invariant open string states contribute:
\begin{equation}
\label{m10}
V_6\int{d^6 k\over (2\pi)^6}\int_0^{\ii\infty} {d\tau'\over\tau'} \, 
{\rm Tr}_{\gamma'}\left\{{\rm e}^{2\pi\ii\tau' (L_0 - a)}\, {\cal P}\right\}~,
\end{equation}
where the trace runs over the adjoint Chan-Paton indices 
and over the Hilbert space generated by the open string
oscillators\footnote{Here $\gamma'=2$ denotes the trace in the R sector, 
$\gamma'=3$ the one in the NS sector and $\gamma'=4$ the trace in the NS sector with
$(-)^F$ inserted.}.
${\cal P}$ denotes the projector $\frac 1 n \sum_{l=0}^{n-1} \hat\omega^l$,
with $\hat\omega$ realizing on the open string states the $\Zbar_n$ generator.
%The generator acts multiplicatively on the open string oscillators
%as required by the geometrical ${\Zbar}_n$ orbifold action
%; for instance
%$(\alpha_n,\bar\alpha)\stackrel{\hat\omega}{to}
%(\omega \alpha_n, \omega^{-1}\bar\alpha)$ for the $X$ and $\bar X$ oscillators.
%and moreover it has the regular action on the Chan-Paton indices discussed
%above.
%Summarizing, the boundary state description of branes at an orbifold 
%singularity is also available in the case of a non-trivial action of 
%the orbifold group on the Chan-Paton factor, requiring only a corresponding
%non-trivially weighed sum over the closed string twisted sectors, as in Eq.
%(\ref{m7}). The description is available for type 0 theories as well; notice
%that in Eq. (\ref{m8}) the contribution from the various parts (R-R or NS-NS,
%with $\eta=\pm$) are considered separately.

From the various ``building blocks'' (\ref{m9}) we construct the 
type II and type 0 expressions 
by taking the combinations required by the definitions (\ref{am1}) and 
(\ref{B+-}), respectively. 
%of the boundary states for type II or type 0  D-branes.
In the type II case, we get 
$\frac 1 2 ({\tilde{\cal A}}^{ij}_3 - {\tilde{\cal A}}^{ij}_4 - {\tilde{\cal
A}}^{ij}_2)^{ab}$. In type 0, we must make the following distinction: for electric-electric
or magnetic-magnetic
interactions ($ab$ or $\bar a\bar b$ indices), where $\eta\eta'=-1$, we get only
contributions from the open string 
NS sector: $\frac 12({\tilde{\cal A}}^{ij}_3 - {\tilde{\cal A}}^{ij}_4 )^{ab}$; 
for electric-magnetic ones, where $\eta\eta'=1$, only  
the R sector ({\it i.e.} worldvolume fermions) contributes: 
$-\frac 12({\tilde{\cal A}}^{ij}_2)^{a\bar b}$. 
Expanding the integrands of these expressions in powers of ${\rm
e}^{\pi\ii\tau'}$ we can count the on-shell states in the world-volume theory 
that carry given Chan-Paton indices, at the various mass levels. 
In particular, the only massless contributions  
arise when $|i-j|$ equals 0 or 1, and the spectrum of the gauge theory
on the world-volume is seen to coincide with the one described in the
previous section.  
%%%%%%%%%%%%%%%%%%%%%%%%%%%%%%%%%%%%%%%%%%%%%%%%%%%%%%%%%%%%%%%%%%%%%%%%%%%%%%%%%%%%%%%%%%%%%%%%%%%%%%%%%%%%
\subsection{Large N conformal invariance}
According to a general argument \cite{nekrasov}, 
large $N$ conformal invariance, whence a vanishing beta function
in the large $N$ limit, is to be expected for the gauge theory on the 
branes at the orbifold singularity. Let us first review the basic
ingredients of this argument. 
Starting from a parent ${\cal N}=4$ theory, one can build 
%theory an orbifold 
a new theory by taking the orbifold 
with respect to  a discrete subgroup $\Gamma$  of the SU(4)$_{\cal R}$ ${\cal
R}$-symmetry group.
%= {\cal T}/\Gamma$, where $\Gamma$ is a discrete subgroup of the SU(4) ${\cal
%R}$-symmetry group. 
If every non-trivial $g \in \Gamma$ acts on a (fundamental) Chan-Paton
index as a {\em traceless} matrix $\gamma_g$,
%. The ${\cal T}'$ field
%theory is then defined by keeping only the $\Gamma$ invariant fields. When the above
%conditions are met, 
the large $N$ vanishing of the  beta function in the orbifold theory can be 
demonstrated to all orders. According to Ref. \cite{nekrasov} this condition on 
the action of $\Gamma$ amounts to imposing that the Chan-Paton indices transform in 
a number of copies of the regular representation of $\Gamma$.

Let us now take the concrete example of $N$ electric and $N$ magnetic type 0 
D3 branes at a $\Cbar^2/\Zbar_n$ singularity. 
Starting from the U($2nN$) ${\cal N}=4$ gauge theory corresponding
to $2nN$ type II branes in flat spaces, the type 0 worldvolume
theory is obtained by orbifolding with $\Zbar_2 \times \Zbar_n$. 
The first factor, the one related to going from type II branes
to type 0 electric plus magnetic branes,
 sits in the centre of SU(4)$_{\cal R}$, 
and its non-trivial generator is represented on the Chan-Paton factor by
${\cal I}$, introduced earlier in this section. 
The second factor is the geometrical orbifold
group and its action on the Chan-Paton factors is taken to be regular, 
as explained at the beginning of this section. 
We are thus taking the orbifold with respect to a discrete subgroup of (a 
SU(2) subgroup of) SU(4)$_{\cal R}$, whose representation on the Chan-Paton factors
meets the above condition.  Thus a large $N$ vanishing beta function is expected.

As a specific check we have evaluated the 2-loop beta function explicitly, taking the field
content of section 4 as basic input. As expected, that spectrum was seen to yield 
a large $N$ vanishing beta function up to two loop order.

%%%%%%%%%%%%%%%%%%%%%%%%%%%%%%%%%%%%%%%%%%%%%%%%%%%%%%%%%%%%%%%%%%%%%%%%%%%%%%%%%%%%%%%%%%%%%%%%%%%
%\section{Conclusions}
%In this paper we have identified the boundary states that describe type 0 D-branes. From the
%relation with their type II counterparts it was then easy to deduce the presence of (chiral)
%fermions on intersections of type 0 D-branes. Furthermore, we have shown the existence of
%anomalous couplings on the branes, either relying on the anomaly inflow argument or by direct
%string computation. 
%
%In a somewhat separate development we have constructed the massless spectrum of the field theory on
%a superposition of electric and magnetic D$3$-branes sitting at an orbifold singularity. The relevant
%cylinder amplitude has also been understood from the boundary state point of view.
%We have illustrated large $N$ conformal invariance of this theory by explicit evaluation of the 
%perturbative beta-function (up to two loops). 
%\medskip
%%%%%%%%%%%%%%%%%%%%%%%%%%%%%%%%%%%%%%%%%%%%%%%%%%%%%%%%%%%%%%%%%%%%%%%%%%%%%%%%%%%%%%%%%%%%%%%%%%%%%%%
%\section*{Acknowledgment}
%This work was supported by the European Commission TMR programme ERBFMRX-CT96-0045.
%%%%%%%%%%%%%%%%%%%%%%%%%%%%%%%%%%%%%%%%%%%%%%%%%%%%%%%%%%%%%%%%%%%%%%%%%%%%%%%%%%%%%%%%%%%%%%%%%%%%%%%%%
\paragraph{Acknowledgment} We thank W. Troost for useful
comments on the manuscript.
%%%%%%%%%%%%%%%%%%%%%%%%%%%%%%%%%%%%%%%%%%%%%%%%%%%%%%%%%%%%%%%%%%%%%%

\end{document}